# Les planétariums numériques - Des musées virtuels pour découvrir la science d'aujourd'hui

Walter Guyot, Hélène Courtois et Jacques Toussaint

**Digital planetariums as virtual museums for discovering today s science**
Abstract :
Our study's concern is about how to enlarge the scope of the virtual museum concept from the mediation activities as they are carried out in today's digital planetariums to training activities dedicated to the public. We assume that these educational and cultural tools must be regarded as real immersive virtual environments and therefore can provide visitors with a new technological framing directly connected with a scientific field in the making, cosmography, in its production environment. One of our goals is to make visitors construct their own knowledge based directly on ongoing research data.

Notre travail souhaite étendre le concept de cybermusée depuis les activités de médiations pratiquées dans les actuels planétariums numériques vers des activités de formation du public. Notre hypothèse est que ces équipements d'éducation et de culture, véritables environnements virtuels immersifs, offrent aux visiteurs un cadre technologique nouveau en lien direct avec un domaine de la science qui s'élabore, la cosmographie, dans son contexte de production. L'un des objectifs que nous affectons à ces outils est de faciliter la construction de connaissances par les visiteurs en prenant appui sur des données de recherche en cours.

Walter Guyot est le responsable scientifique du planétarium de Vaulx-en-Velin (Lyon – France) depuis dix ans. Il participe majoritairement à la programmation culturelle et éducative de cet équipement destiné à la transmission de connaissances en direction de tous les publics. Il complète présentement sa formation à l'Université Lyon 1 en rédigeant une thèse de doctorat en didactique de la cosmographie et de la cosmologie. – w.guyot@ipnl.in2p3.fr

Hélène M. Courtois est docteure en astrophysique et maître de conférences à l'Université Lyon 1 (France) depuis 1997. Elle a dirigé l'équipe de cosmologie de l'observatoire de Lyon de 2002 à 2006. Depuis 2013 elle dirige le groupe de recherche Cosmologie observationnelle/Euclid de l'Institut de physique nucléaire de Lyon. Elle est experte scientifique auprès du planétarium de Vaulx-en-Velin depuis 2005. – h.courtois@ipnl.in2p3.fr

Après une thèse sur la nucléosynthèse (pilotée par H. Reeves), Jacques Toussaint s'est mobilisé sur les aspects de la formation à la médiation (enseignement et musées) en didactique de la physique. Intervenant puis responsable de plusieurs masters, directeur de nombreuses thèses, ses thématiques de recherche ont porté sur les difficultés d'apprentissage des concepts scien- tifiques qu'expliquent leurs constructions historiques et leur épistémologie. Il est aujourd'hui professeur émérite à l'Université Lyon 1. – jacques.toussaint@univ-lyon1.fr

Les progrès incroyablement rapides de la technologie informatique ont mis à la disposition de nos planétariums contemporains de nouveaux dispositifs numériques de réalité virtuelle[1] pour cultiver et former les publics à la cosmographie[2] moderne. Plus qu'une simple (r)évolution, ces équipements d'éducation et de culture sont devenus de véritables environnements virtuels en 3D, passant de la représentation d'un monde clos à celle d'un Univers infini. À la manière d'un observatoire virtuel[3], ces planétariums ont acquis la possibilité de conserver et de mettre à la disposition de leurs publics des catalogues d'observations classées, archivées ou en cours d'analyses par les chercheurs. Les enjeux culturels aux- quels peuvent prétendre ces institutions sont triples : le premier est de devenir de véritables musées virtuels collectionnant des données d'observations astronomiques sur lesquelles la science investit un travail de recherche, le deuxième est de montrer la science « telle qu'elle se fait[4] » et, enfin, le troisième est de faciliter la construction de connaissances par les visiteurs en prenant appui sur des don- nées de recherches en cours. Le stockage et la conservation de données observationnelles qui n'existent que sous forme numérique offrent en effet l'occasion aux planétariums numériques de garder une trace du processus vivant d'un savoir-faire propre à une communauté de chercheurs ; cela permet également d'envisager des situations de médiation des plus insolites pour transmettre au public des savoirs en cours d'émergence. Nous formulons l'hypothèse que ces structures, véritables environnements virtuels immersifs, proposent un cadre techno- logique nouveau en lien direct avec un domaine de la science qui s'élabore, la cosmographie, dans son contexte de production. L'idée de montrer une science en action se trouvait déjà dans les aspirations du Palais de la découverte[5] à ses origines et s'inscrivait, avant l'heure, dans une approche que proposera Bruno Latour[6] ou le courant Public Understanding of Research[7] dans les années 1990. Comment les planétariums numériques montrent-ils la science en action ? Quelle forme de médiation ces outils offrent-ils à leurs visiteurs pour les initier à des savoirs scientifiques émergents ? Dans cet article, nous décrivons comment notre recherche en cours conduit à valider notre hypothèse sur la réussite d'apprentis- sages des visiteurs d'un planétarium, sur un objet scientifique encore en discussion, le rôle de la « matière noire » dans l'évolution de l'Univers à grande échelle.

**L'ambition muséale des planétariums**

Le Conseil international des musées (ICOM)[8] reconnaît depuis 1989 les centres de culture scientifiques ainsi que les planétariums comme répondant à sa définition du musée[9]. Le XX[e] siècle a vu en effet se créer de nouveaux établissements à vocation scientifique n'ayant pas (ou peu) de collections tels que le Palais de la découverte et la Cité des sciences à Paris, en France, ou encore l'Exploratorium à San Francisco, aux États-Unis. Toutes ces institutions constituent des organismes à but non lucratif ayant pour point commun d'être des lieux d'éducation et de diffusion consacrés à la culture, au patrimoine et à la science en direction du grand public[10]. Pourtant, si la définition de l'ICOM a cherché à réunir tout ce qu'on entend habituellement par le terme de musée, elle demeure essentiellement cadrée sur l'acquisition, la conservation, l'étude et la mise en valeur de collections[11]. À partir de cette définition, nous pouvons réserver, à la manière de Michel Van Praët (1989)[12] ou de Marine Soichot (2011)[13], le terme de « musée » aux établissements dont une majeure partie de l'activité est consacrée aux collections et celui de « centre de sciences » – entendu comme centre de culture scientifique et technique – aux établissements de médiation scientifique accueillant des publics, sans toutefois avoir d'activité majeure de collections et de conservation. Réévaluée en permanence, la définition de l'ICOM a également été complétée en 2007[14] par les notions de « patrimoine[15] » et « d'immatériel », permettant ainsi aux institutions muséales d'élargir le champ autour de l'objet matériel qu'elles conservent[16], mais aussi de s'intéresser aux possibilités offertes par la mémoire numérique pour leurs collections. L'idée d'un patrimoine numérique n'est pas nouvelle et s'est construite progressivement depuis la fin de la Deuxième Guerre mondiale[17]. Aujourd'hui, la définition donnée à ce patrimoine d'après la charte de l'Organisation des Nations Unies pour l'éducation, la science et la culture (UNESCO) sur la convention du patrimoine numérique[18] se résume comme étant constitué soit par la numérisation de documents historiques ou patrimoniaux existants sur un autre support (papiers, livres, objets techniques anciens...), soit par la sauvegarde de documents créés originellement dans un format numérique[19]. Ce patrimoine a ainsi intéressé un musée d'un nouveau genre, qualifié de virtuel, utilisant Internet et les nouvelles technologies à des fins de promotion, d'éducation, ou de conservation de collections d'objets numérisés[20]. Le concept de musée virtuel s'est répandu à compter des années 1990 et a trouvé lui aussi une légitimité dans la définition de l'ICOM[21]. Dans un document consacré à une approche du concept de musée virtuel, Werner Schweibenz[22] le définit comme « une collection logiquement agencée d'objets numérisés et composée de divers supports média qui, par sa connectivité et son caractère multi-accès, permet de dépasser les modes traditionnels de communication et d'interaction avec le visiteur ». Sa définition fait suite à celle donnée par David Bearman[23] qui englobe le musée virtuel dans la catégorie du « musée hors les murs ». Selon Schweibenz[24], l'étendue de la virtualité ou la nature du musée en ligne n'a pas d'importance, car il faut avant tout offrir une intégration enrichie d'objets d'apprentissage et d'information à leur sujet[25]. Toutefois, la définition d'un musée virtuel demeure en constante évolution et une grande diversité de termes sont utilisés de manière interchangeable pour le désigner, entre autres : musée électronique, musée numérique, musée en ligne, musée hypermédias, méta-musée[26]. Ce musée d'un nouveau genre invite donc à une réflexion autour d'un musée « hors les murs » où les collections et les objets sont désormais virtuels et numériques, réunis au sein de banques de données en vue de leur conservation et de leur présentation. L'évolution des musées virtuels redessine le paysage muséal, libérant l'accès aux collections au-delà de toutes barrières institutionnelles. Une nouvelle forme de médiation des œuvres et des savoirs se profile ainsi à travers le réseau – le Web[27]. Dans la lignée du musée virtuel, des institutions ont également développé un accès à des collections de données sur Internet à partir de logiciels spécifiques dans le but, cette fois, de permettre à des chercheurs ou à des amateurs de travailler à distance sur des projets de recherche. Elles ont ainsi partagé avec la communauté des données issues de la recherche qui n'existent que sous forme numérique[28]. Ces données, collectées, archivées, conservées et diffusées de manière judicieuse dans un objectif de recherche, constituent un patrimoine numérique scientifique à part entière. Un observatoire virtuel précisément né de cette réunion de plusieurs bases de don- nées spécialisées (images, spectres, données numériques...) permet aux astronomes de croiser leurs données observationnelles. Notre étude s'intéresse à cette composante « née numérique[29] » qui résulte d'un processus de production initiale « tout-numérique ». En effet, la communication en direction du public de ces données de la recherche nécessite parfois un mode d'accès spécifique et contextualisé à l'aide d'environnements nouvellement dédiés à leur représentation. De quoi s'agit-il et en quoi cela intéresse-t-il les planétariums ?

Le premier domaine dans lequel s'est massivement déployé le numérique depuis les années 1950 a été celui de la science et de ses applications dans le militaire et le spatial[30]. Une ère de profonde mutation a fait émerger un nouveau champ de mémoire pour l'humanité : celui des données élaborées par les expériences de laboratoires ou les observations collectées par de multiples dispositifs d'enregistrement (satellites, accélérateurs de particules, télescopes, sondes spatiales, micro-caméras, etc.). Cela constitue une somme de données toujours plus importante à traiter et à gérer par une informatique toujours plus puissante – parfois plusieurs centaines de gigaoctets par jour. Certaines de ces données font figure de véritables événements historiques qui ne pourront plus jamais être renouvelés dans le temps (photos des premiers pas de l'homme sur la lune, observation filmée de la terre depuis l'espace, explosion de supernova...) D'autres sont capitales d'un point de vue scientifique, puisqu'elles permettent d'aboutir soit à de nouvelles découvertes, soit à la confirmation ou non de modèles ou d'hypothèses scientifiques. Dans le même temps, de nouveaux outils de visualisation et de représentation de ces données, tels les logiciels de simulation ou de production d'images de synthèse en 2D et 3D ont fait leur apparition. Ils ont offert de nouvelles manières d'accéder à ce patrimoine « né numérique » dans un objectif de manipulation, de recherche, de mise en valeur ou d'éducation31. Les planétariums, armés de leurs nouvelles technologies, se sont intéressés à ces résultats de la recherche passée et actuelle comme témoins d'une science en action. Un nouveau regard, une nouvelle façon de voir le monde s'est donc élaborée au cœur de ces institutions, les rapprochant des notions de patrimoine, de conservation et de collection32. Dans ces structures, l'Univers se montre désormais à travers des modèles numériques à partir desquels le public peut acquérir non seulement des connaissances, mais aussi de véritables compétences en interagissant seul ou avec un médiateur dans l'espace virtuel proposé. Une aventure nouvelle pour le public qui découvre, à partir de données scientifiques, la démarche scientifique sous-jacente qui les élabore, mais également des valeurs, des émotions autour de savoirs en jeu. Aujourd'hui, les planétariums numériques ne permettent donc pas seulement d'accumuler des témoins sur les- quels la science investit un travail de recherche, ils offrent aussi la possibilité de garder la trace de ce que nous faisons de ces témoins. Dans cette optique, les planétariums peuvent proposer à leurs publics de prendre du recul sur ces témoins de la recherche « nés numériques » en apportant une réflexion critique grâce à une présence humaine. En effet, nous pensons que l'apport du numérique a permis de développer des activités de médiation dans lesquelles un médiateur peut relier au sens fort la recherche scientifique aux données sur lesquelles elle se base (données expérimentales, données d'enquête, corpus, etc.). Dans cette perspective, le public a donc la possibilité de vérifier la pertinence et la véracité de cette recherche, de ce qu'elle fait des données de départ qu'elle emploie, convoque et interprète au quotidien.
Nous rejoignons ici Bruno Latour33 sur l'idée de « démocratie scientifique34 » ou encore Benoit Habert35 sur celle d'une mémoire critique, qui s'inscrivent pleinement dans cette logique. L'ambition que nous attribuons aux planétariums numériques – et plus particulièrement dans notre cas spécifique d'utilisation du planétarium lyonnais – est bien d'offrir à la population un contexte nouveau pour mieux comprendre les enjeux et les intérêts de la science moderne ; l'originalité qu'apporte leur technologie est de pouvoir utiliser les données sur lesquelles la science investit un travail de recherche pour éveiller le public à des savoirs scientifiques émergents. La révolution numérique de ces équipements apporte donc au citoyen la possibilité de replacer des données de recherche dans leur contexte de production et de s'initier au cadre théorique qui les a engendrées, mais aussi aux interprétations qu'elles suscitent.

**Les planétariums numériques: des musées virtuels en effervescence**

Objet de médiation par excellence, un planétarium est une installation permettant de représenter les mouvements des astres sur une voûte hémisphérique, grâce à des projections lumineuses. L'être humain se forge des connaissances principalement en analysant le monde et en élaborant des représentations de ce monde par des processus cognitifs. C'est donc bien en visant des apprentissages humains que se sont développés les planétariums, afin que l'Homme comprenne le monde et qu'il se construise une idée de l'Univers dans lequel il se trouve. Les planétariums se donnent ainsi pour objectif de représenter un monde ressemblant à la perception admise par une communauté scientifique de notre réalité tout en offrant une simulation réaliste des phénomènes célestes observés. L'astronomie étant essentiellement une science d'observation et de modélisation, l'image de cette science véhiculée par les planétariums est ainsi respectée. Pour cela, ces outils ont bénéficié, à chaque étape de leur évolution technique, de la vision et des interrogations de l'homme sur le cosmos. Quelle que soit l'époque, les planétariums ont en effet toujours été des marqueurs de la représentation de l'Univers issue d'un consensus largement admis, d'une culture ou d'un savoir partagé.

Historiquement, le terme « planétarium » pro- vient de « planétaire », une maquette représentant initialement le monde clos géocentrique où s'animent les « errants », c'est-à-dire les planètes visibles, et les deux luminaires, le soleil et la lune. Il s'agissait d'une première tentative pour expliquer le monde, un objet artisanal et instrumental à vocation cosmographique pour expliquer, modéliser, simuler, représenter et mesurer l'organisation du cosmos. Le premier d'entre eux est attribué à Archimède de Syracuse (-287– -212) et leur concept évoluera indépendamment de la confection de globes célestes de type sphère armillaire, réservés uniquement au repérage des astres et à la cartographie du ciel36. Cependant, comme le soulignait Michel Dumont37, il faut attendre la fin du XVIIe siècle pour qu'apparaissent les premiers planétaires au sens moderne, c'est-à-dire capables de reproduire mécaniquement et avec précision le mouvement des planètes. En 1862, Christian Huygens (1629–1695) rendit les maquettes tributaires d'un mécanisme d'horlogerie ; les vitesses de déplacement étaient alors obtenues par l'intermédiaire de multiples roues dentées. Au XIXe siècle, le professeur Roger Long fait construire un stellarium qui représente un ciel étoilé pouvant accueillir une trentaine de personnes38. En 1912, le globe de Wallace Walter Atwood préfigurera l'ingénieuse idée de la société Carl Zeiss d'Iéna (Allemagne) qui fusionnera, quelques années plus tard, les modèles de planétaire et de globe pour créer le premier planétarium moderne. Cette innovation sera à jamais liée à Oskar Von Miller (1855–1934), fondateur du Deutsche Museum de Munich, et à Walther Bauersfeld (1879– 1959) de la société Zeiss. Le planétarium moderne est alors inventé et la description du ciel va trouver une forme de sacralisation lors de présentations audacieuses grâce au simulateur opto-mécanique et aux récits des conférenciers sur l'univers des éphémérides. Au-delà de l'émotion et de l'admiration que procura le miracle technologique Zeiss, les planétariums poursuivront leur mutation pour devenir, dès 1965, de véritables théâtres multi-images, davantage en harmonie avec la grande époque de la conquête de la lune et l'intérêt politique pour des communications de la science plus prestigieuses en direction du grand public39. Le multimédia s'est imposé vers 1990, aboutissant sur toute la planète au développement des planétariums numériques du troisième millénaire. Cette métamorphose fut opérée par la société américaine Evans et Sutherland dès 1983, grâce au premier prototype de planétaire graphique, le Digistar I, installé à Salt Lake City (Utah, É.-U.). Ce type de système consiste à diffuser une vidéo en pleine voûte (full dome) sur l'écran hémisphérique du planétarium, soit à l'aide d'un projecteur unique équipé d'une optique de type fish-eye (projection à 180 degrés), soit grâce à une combinaison de plusieurs projecteurs recouvrant chacun une partie de la voûte. Cette évolution a permis aux planétariums de devenir de véritables environnements virtuels d'apprentissage humain (EVAH) qui utilisent désormais la réalité virtuelle pour transmettre des savoirs ou des pratiques scientifiques auprès du grand public. L'Univers présenté dans ces environnements virtuels en 3D est toujours très proche de la perception commune de notre réalité, mais il est aussi le reflet de modèles scientifiques conjugués à l'emploi de simulations numériques ; les planétariums numériques placent donc les visiteurs non pas devant des objets matériels comme le font les expériences réelles, mais bien devant des représentations symboliques censées rendre compte de ces objets. Comme nous l'avons déjà évoqué, l'originalité de ces structures est de prendre appui sur des données scientifiques « nées numériques », c'est-à-dire des archives d'observations astronomiques, afin de permettre à leurs publics de construire leurs propres connaissances sur des savoirs en cours d'émergence. Mais quelles innovations la réalité virtuelle apporte-t-elle aux planétariums numériques pour y arriver ?

La réalité virtuelle est un mode de représentation intégrant l'ensemble des techniques de rendu, d'interaction et d'immersion[40]. Elle offre la possibilité de générer des environnements virtuels qui placent un sujet humain en situation d'interaction mais également d'immersion dans un univers recréé de toutes pièces. Le sujet humain a alors la possibilité d'éprouver une expérience de la réalité dans un univers virtuel en 3D. Jacques Tisseau caractérise la réalité virtuelle comme « un univers de modèles au sein duquel tout se passe comme si les modèles étaient réels parce qu'ils proposent la triple médiation des sens, de l'action et de l'esprit ».[41] Cet auteur montre que l'homme appréhende le réel par l'entremise de ces trois médiations. Un système de réalité virtuelle se distingue donc d'une autre application informatique par le fait qu'il offre à l'utilisateur la sensation d'être dans un monde virtuel et d'y agir[42]. L'utilisateur doit aussi avoir « le sentiment authentique d'exister dans un monde autre que le monde physique où le corps se trouve »[43], c'est-à-dire un sentiment de présence. Cette notion de présence de l'utilisateur dans un environnement virtuel se décline selon deux composantes : l'immersion[44], pouvant être multisensorielle, et l'interaction[45]. Mais pour être complet, un environnement virtuel doit également être un lieu où les objets se comportent de manière autonome sans aucune action de l'utilisateur. Ainsi, Jacques Tisseau et Fabrice Harrouet[46] ont envisagé trois axes pour caractériser un système de réalité virtuelle : autonomie, interaction et immersion. Ce repère fournit un cadre pour positionner tout système informa- tique vis-à-vis de l'utilisation qui peut être faite de la réalité virtuelle (ill. 1). Chaque sommet du cube théorique illustre ici un cas « idéal ».

La simulation informatique et plus particulièrement la réalité virtuelle ont trouvé des applications dans des contextes d'apprentissage qui nécessitent le transfert de compétences et la mise en situation des apprenants. Elles offrent alors la possibilité de reproduire des expériences coûteuses (d'un point de vue maté- riel), risquées (d'un point de vue humain) ou inaccessibles (la représentation d'aspects non visibles du réel ou impossibles à représenter dans leur forme réelle) pour l'homme[47]. Les planétariums ont toujours fait appel à de telles simulations comme substitut d'expériences réelles. Le but de ces systèmes informatiques dédiés à la formation de l'Homme est donc de faire acquérir à des apprenants des connaissances ou des gestes techniques définis par des objectifs pédagogiques[48]. Grâce à cette technologie, les planétariums numériques (ill. 2) sont désormais des environnements virtuels qui offrent à leurs visiteurs la possibilité de passer d'un monde clos à un univers infini et en évolution. Le visiteur évolue alors dans un espace de représentation réaliste, immersif, tridimensionnel, calculé en temps réel[49]. La projection d'images et de son spatialisés sur une voûte hémisphérique renforce la sensation du visiteur d'être immergé dans un espace réel. Le public n'est alors plus en contact direct avec des objets réels, mais bien avec leur représentation symbolique, mise en scène au cœur de nouveaux espaces, virtuels eux aussi. Dès lors, il ne s'agit plus de simples copies du réel, mais de substituts, de modèles de la réalité.

Le visiteur se trouve donc plongé dans un monde reconstruit de toutes pièces dans lequel il lui est possible de s'affranchir des contraintes de temps, d'échelle ou d'espace ou encore de découvrir des endroits difficilement accessibles (la planète Mars, la ceinture d'astéroïdes, les amas de galaxies ou l'infiniment petit...). Cette technologie est désormais essentielle pour mettre à la disposition du public des bases de données donnant accès à des observations passées, classées, archivées et menées par différents programmes de recherche, mais aussi les résultats obtenus par les grands observatoires, leurs programmes majeurs d'atlas et d'observations de régions précises, les missions spatiales... Cette visualisation de collections de données numériques sous forme de modèles numériques – témoins d'une science en action et de la description de l'Univers par la science à une époque donnée – peut être à la source d'un certain nombre de situations de médiation au cours desquelles sera privilégiée l'interaction temps-réel entre le public, le médiateur et les scènes présentées. La réalité virtuelle a apporté aux planétariums numériques la capacité de se redéfinir comme de véritables musées virtuels collectionnant des données scientifiques, seules capables de montrer l'élaboration actuelle de la recherche dans les domaines de la cosmographie et de la cosmologie. Le caractère innovant des planétariums numériques nous a conduits à traiter scientifiquement l'hypothèse d'un apprentissage efficace des spectateurs. Ces équipements sont aujourd'hui basés sur des technologies permettant de positionner les visiteurs en tant qu'acteurs de leur formation à la science « telle qu'elle se fait ».

**Des planétariums – musées pour découvrir « la science telle qu'elle se fait »**

Montrer une science en action, c'est présenter la science la plus actuelle, celle qui n'est pas encore « normalisée » au sens des épistémologues. C'est une approche similaire que proposera le courant Public Understanding of Research dans les années 1990[50], ouvrant ainsi la voie à des activités de médiation dans lesquelles le public découvrira les derniers résultats de la recherche en cours. Mais comment les planétariums numériques peuvent-ils montrer la science telle qu'elle se fait ?

Pour cela, nous nous sommes intéressés à une séquence en temps réel du planétarium de Vaulx-en-Velin[51] qui propose depuis plus de dix ans le projet « Un ciel de galaxies »[52] aux adultes et aux adolescents. Celui-ci vise à initier le public à la cartographie et à la géographie de l'Univers à grande échelle à partir de catalogues d'observations de galaxies. Depuis près de trente ans, des projets observationnels de grande ampleur ont en effet permis de progresser de façon remarquable sur la répartition de la matière jusqu'à des distances de l'ordre de plusieurs milliards d'années- lumière. La richesse et la profondeur des relevés systématiques du ciel ont ainsi apporté une image renouvelée de l'Univers qui nous entoure. D'une part, la mesure de la vitesse de récession d'un grand nombre de galaxies, fournissant par l'intermédiaire de la loi de Hubble[53] l'accès à la troisième dimension, a modifié la vision de l'Univers issue de catalogues projetés autrefois en 2D. Les nouvelles données d'observations ont révélé une structuration cellulaire de la matière à grande échelle. L'image que nous avions de notre univers au début du $XX_e$ siècle a donc changé. Désormais l'Univers observable est perçu comme une alternance de grands vides, de murs et de filaments dont les propriétés topologiques s'apparentent à celles d'une « éponge », tandis qu'aux échelles extragalactiques plus petites, les galaxies sont les objets astronomiques les plus discernables et représentent une surdensité de matière « lumineuse » organisée en amas et superamas. D'autre part, la matière dont nous sommes constitués (baryons) n'est qu'une faible part de celle qui dominerait la dynamique des amas ; la présence de matière noire gravitationnelle- ment attractive serait en effet une composante majeure de l'Univers après « l'énergie noire » qui serait répulsive[54]. Par ailleurs, l'étude de l'évolution spatio-temporelle des structures grâce aux mesures des vitesses particulières de leurs galaxies a ouvert la voie à une cartographie 4D de la distribution observée en apportant des informations précieuses sur la répartition et la quantité de cette matière noire (matière ordinaire baryonique, neutrinos, particules « exotiques »). La formation et l'évolution des structures dépendent de la densité moyenne de l'Univers qui elle-même gouverne la géométrie et le destin de ce der- nier. Cette nouvelle image « dynamique » de la distribution et du contenu de la matière de notre Univers suscite un regain d'intérêt dans la recherche pour la cartographie des grandes structures de galaxies. Elle offre ainsi un contexte observationnel unique pour tester les scénarios théoriques de notre cosmologie moderne. C'est également une entreprise où les savoirs encore incertains sont susceptibles d'intéresser le public et de leur offrir une première initiation à la cosmologie moderne. Depuis 2012, ce planétarium propose un catalogue de galaxies issu du projet Cosmic Flows, du groupe Cosmologie observationnelle/Euclid à l'Institut de physique nucléaire de Lyon (IPNL). Ce projet de recherche a pour objectif de cartographier les densités de matière noire et d'énergie noire locales dans un rayon de 100 mégaparsecs (Mpc)[55] autour de notre galaxie[56]. La méthode retenue par les chercheurs est celle de la mesure des distances des galaxies. En effet, lorsque l'on est en pos- session à la fois de la distance et de la vitesse d'une galaxie dans l'expansion générale, on a alors accès à sa vitesse particulière : sa vitesse due à l'environnement gravitationnel de cette galaxie. Ces savoirs émergents ne sont pas encore stabilisés et sont représentatifs d'une recherche « telle qu'elle se fait ». L'originalité pour ce planétarium de s'associer à ce pro- gramme de recherche est double : 1) permettre de renouveler constamment ses bases de don- nées d'objets astronomiques (catalogues de galaxies) et ainsi placer ses visiteurs en contact étroit avec les résultats de la recherche la plus actuelle ; 2) présenter des savoirs en cours d'émergence dans leur contexte de production. L'ambition affichée ici n'est donc pas de reproduire des expériences déjà faites ou de mettre en scène uniquement des méthodes scientifiques éprouvées. Il ne s'agit pas non plus de
réactiver le principe du deficit model[57], mais plutôt de proposer de sortir d'une communication unidirectionnelle du savant à destination de visiteurs novices. Comment le public peut-il découvrir et participer à la science « telle qu'elle se fait»?

Depuis une observation terrestre d'un ciel de galaxies simulé (ill. 3) sur la voûte immersive du planétarium, le public est convié à un voyage virtuel dans un Univers reconstitué en 3D (ill. 4). Nous avons, à partir du catalogue Cosmic Flows, dressé un fichier en coordonnées cartésiennes X Y Z dans un référentiel galactique répertoriant près de 10 000 galaxies. Nous avons ainsi généré un nuage de points de galaxies centré sur la Voie lactée de 100 Mpc permettant d'apprécier leur distribution spatiale. Nous avons pu projeter et manipuler cet échantillon sous la forme d'un objet 3D grâce au planétaire Digistar 3[58]. L'objectif premier était de pouvoir reproduire des observations virtuelles à partir de données réelles issues de campagnes d'observations astronomiques. Il s'agissait alors de décrire avec le public la répartition des galaxies en amas et en filaments autour de grandes bulles vides, tout comme le font les astronomes d'aujourd'hui, mais à par- tir de la simulation des mouvements observés et mesurés des galaxies. Les visiteurs orientent leur position dans l'espace virtuel du catalogue de galaxies à l'aide du médiateur et découvrent le rôle de la gravitation dans la distribution de la matière à grande échelle. L'originalité de cette expérience réalisée avec le public est donc de pouvoir intégrer les résultats quotidiens du projet Cosmic Flows et de montrer la répartition spatiale des vitesses particulières des galaxies sur l'ensemble d'une couverture 3D (ill. 5). La mise en situation des visiteurs dans cet environnement virtuel leur permet de s'immiscer directement au cœur de l'interprétation scientifique dont le support de médiation est l'ensemble du catalogue avec ses données numériques en représentation. Les visiteurs peuvent ainsi reconstruire par eux-mêmes la cartographie d'ensemble de l'Univers local à partir des mouvements qui l'animent : c'est le sens du travail de recherche développé aujourd'hui par Hélène Courtois et ses collaborateurs et que nous nommons « cosmographie moderne » ou « cosmographie dynamique ».

Les visiteurs sont ainsi encouragés par le médiateur à s'interroger sur les observations perçues, à décrire les phénomènes simulés ou à proposer leurs propres interprétations. La simulation doit leur permettre d'observer tout d'abord le déplacement de l'expansion de la paroi d'un grand vide sur le bord duquel notre galaxie se trouve, puis la chute du groupe local de notre galaxie vers un très gros amas de galaxies proches : l'amas Virgo. Avec l'amas Virgo, le dispositif de la présentation doit permettre aux visiteurs de repérer une autre chute à très grande vitesse, dans une direction bien précise ; il s'agit de la position possible du « grand attracteur » (ill. 6). Une région qui a fait couler quantité d'encre depuis quinze ans dans la recherche en cosmographie, car mystérieusement elle ne comporte pas de superamas de galaxies pouvant expliquer cette forte attractivité gravitationnelle. Dans les années 1980, cette structure a intéressé particulièrement les astronomes qui ont mis en évidence que notre groupe local s'écartait du mouvement d'expansion général de l'Univers à la vitesse de 366 ± 125 kilomètres par seconde dans la direction du Centaure. D'où l'idée qu'une énorme quantité de matière pouvait attirer notre galaxie et les galaxies environnantes. Il n'existe à ce jour aucune réponse tranchée sur le sujet, mais seulement plusieurs hypothèses en cours de publication : superamas de matière noire, effet de construction vectorielle. La recherche du grand attracteur a ainsi occupé de nombreuses équipes d'astronomes comme celles de Dominique Proust, de Georges Paturel ou d'Hélène Courtois en France. L'amas Shapley 8 fut un bon candidat jusqu'à très récemment. Le programme de recherche Cosmic Flows d'Hélène Courtois propose aujourd'hui une tout autre méthodologie : c'est un processus dans lequel le public peut s'immiscer pour comprendre une recherche dont tous les résultats ne sont pas encore publiés et stabilisés. C'est pourquoi, d'ailleurs, sa contribution de chercheuse est nécessaire de manière constante dans l'élaboration du projet « Un ciel de galaxies » en permettant de former directe- ment les médiateurs à l'analyse de ces catalogues et à l'actualité de sa recherche. Ce parrainage[59] du scientifique est la condition sine qua non pour amener le médiateur avec son public à reconstruire, dans une approche scientifique, l'évolution de l'Univers local à l'aide de la position et du mouvement propre des galaxies, hors expansion. Le projet « Un ciel de galaxies » permet ainsi pleinement d'engager les visiteurs dans un processus de modélisation pour cartographier la matière à grande échelle et intégrer la notion de matière noire. L'univers scientifique est ainsi mis à la portée du public pour lui offrir l'occasion avant les autres de découvrir, par l'action, des savoirs en cours d'élaboration.

**Vers un modèle de la situation médiatique dans les planétariums numériques**

La fréquentation de ces structures de culture scientifique et technique pose la question de la transmission et de l'appropriation de connaissances spécifiques dans un contexte qui n'est pas celui de l'école, ce qui nous place aussi dans un contexte connu des recherches en didactique des sciences. La spécificité de l'apprentissage dans un planétarium pourrait en effet être qualifiée de « non formelle » : le type d'apprentissage, les objectifs, la situation diffèrent complètement du cadre scolaire. Dans ces conditions, le visiteur découvre le lieu selon ses envies avec des intentions à la fois culturelles et de loisir. Lors d'une séance de planétarium, les activités proposées sont diverses et font appel à l'utilisation de nombreux médias, telle la projection de films pré-calculés ou de séquences d'animations thématiques en temps réel avec un médiateur. Une séance est ainsi une véritable mise en scène médiatique que nous définissons comme une situation d'apprentissage culturel basée sur le partage des connaissances, au cours de laquelle sont mélangés les acquis intellectuels (comprendre, questionner, analyser), les actions physiques et intellectuelles (agir, jouer, sélectionner...) et la dimension socio-affective (sensation, émotion, curiosité, envie, plaisir...). Le visiteur vit pleinement une expérience de visite dans un monde qu'il affronte avec ses propres conceptions et schèmes intériorisés. Or, les situations médiatiques proposées dans ces planétariums ne sont pas uniquement le fruit d'une transposition didactique de savoirs savants, de l'avis d'Yves Chevallard[60], ou de pratiques professionnelles, selon Philippe Perrenoud[61]. Elles résultent d'une nouvelle forme de transposition qui prend en compte l'ensemble des processus cognitifs élaborés pour transmettre des savoirs et des pratiques scientifiques dans un contexte culturel : cela renvoie au concept de transposition médiatique décrit par Jack Guichard et Jean-Louis Martinand[62]. Le temps réel et de la réalité virtuelle justifient par ailleurs l'emploi de la pédagogie constructiviste dans les planétariums, pédagogie qui considère l'apprenant comme acteur de ses apprentissages. Un rapprochement cognitif s'est donc opéré entre les différents acteurs de la situation médiatique proposée. Le médiateur a ainsi un rôle de compagnonnage cognitif au cœur de l'EVAH exploré par le public : il pose des questions ouvertes, participe à l'activité, observe tout en laissant des initiatives aux visiteurs, relance ou oriente le débat et les questions du public. Cela rejoint la question des relations des interactions sociales comme fondement de l'apprentissage[63] ou du rôle des pairs en tant que « facilitateurs » de l'apprentissage déjà proposé par Lev S. Vygotski dans les années 1930[64] ; on peut parler alors d'un mode « socio-constructiviste » des apprentis- sages. Le rôle fondamental du médiateur dans l'action éducative de tout musée est en effet central, comme le souligne Brigitte Zana[65] :
« Cependant force est de constater que le meilleur objet muséologique, la meilleure exposition ont leurs limites et ne pourront suffire à l'apprentissage [...] qui nécessite l'intervention humaine ! » Les planétariums numériques en tant que musées virtuels se différencient ainsi des autres musées de la toile par le fait qu'ils sont des lieux habités par une médiation humaine, seule capable pour nous de faciliter une construction cohérente de connaissances par le public. En effet, c'est grâce au médiateur que les données numériques collectées et présentées trouvent du sens auprès du visiteur. Pour comprendre et montrer comment les planétariums numériques permettent à leur public de construire des connaissances sur des données de recherche en cours, nous avons adopté une approche théorique suivant une double réflexion. La première, qui s'appuie sur les travaux de Lev S. Vygotski et de Jérôme S. Bruner[66], pose la présence du médiateur comme étant nécessaire pour mettre en récit, avec l'environnement numérique de planétarium, la science telle qu'elle se fait, c'est-à-dire de formuler avec le public des hypothèses, une démarche d'investigation scientifique, des vérifications ou des incertitudes en lien avec des données « nées numériques », dans notre cas, issues de la cosmographie dynamique. La seconde réflexion, s'appuyant sur l'approche instrumentale de Pierre Rabardel[67], conduit à dissocier l'environnement numérique du planétarium d'une part en tant qu'instrument et d'autre part en tant que médiateur du contenu d'un savoir. Sur cette double base, nous pro- posons de présenter le travail du médiateur
dans un modèle de « tétraèdre médiatique » adapté du carré médiatique de François-Xavier Bernard[68] rendant compte des caractéristiques de l'interaction qui se joue entre les quatre composantes de la situation médiatique immersive observée : 1) l'environnement virtuel, 2) le médiateur, 3) le public-visiteur, 4) le savoir et les pratiques scientifiques visés. Afin de mettre à l'épreuve notre hypothèse sur l'impact éducatif de la situation médiatique proposée lors d'une séance de planétarium sur les visiteurs, nous avons mis en place une méthodologie d'enquête se fondant sur des questionnaires pré-test et post-test ainsi que sur des entretiens individuels semi-directifs appuyés par un guide d'entretien préalable- ment établi. Cette phase du travail n'est pas encore aboutie. Notre ambition dans cette étude est de permettre d'interpréter (à l'instar d'une grille de lecture) grâce au tétraèdre l'évolution des interactions médiatiques opérées lors d'une séance de planétarium et ainsi de mieux comprendre comment se transmettent, auprès du public, des savoirs issus de la science actuelle.

**Conclusion**

L'originalité des planétariums numériques est de pouvoir proposer aux publics des environnements uniques où sont montrées dans leur contexte de production des données de la recherche qui n'existent que sous forme numérique. Ce patrimoine numérique de la recherche scientifique passée mais aussi actuelle permet à ces lieux d'en conserver une trace sous forme de modèle numérique, tout en présentant la démarche scientifique dans laquelle ces données ont été générées. C'est bien une aventure sensorimotrice et cognitive dans un Univers de modèles que proposent à leurs visiteurs les nouveaux planétariums numériques. Un monde virtuel, mais bien réel, où il est possible d'investir un travail de recherche, de formuler des hypothèses, de vérifier des possibles ou de mettre en place une démarche d'investigation scientifique. Cela confère à ces équipements un statut de véritables musées virtuels collectionnant des données numériques sur lesquelles la science investit un travail de recherche, mais également de devenir des lieux où la science se montre « telle qu'elle se fait » où le médiateur devient le catalyseur de l'apprentissage du public. C'est pour nous une image nouvelle, active, du « musée scientifique ».

1983.
64   WEIL-BARAIS, Annick et Marcela RESTA- SCHWEITZER. « Approche cognitive et développementale de la médiation en contexte d'enseignement-apprentissage ». La nouvelle revue de l'adaptation et de la scolarisation, n₀ 42, 2ᵉ trimestre, 2008, p. 83–98.
65   ZANA, « Histoire des musées, médiateurs et éducation scientifique », op. cit., p. 5.
66   Id. : et BRUNER, Le développement de l'enfant..., op. cit.
67   RABARDEL, Pierre. Les hommes et les technologies : une approche cognitive des instruments contemporains. Paris : Armand Colin, coll. « U : Série Psychologie », 1995.
68   BERNARD, François-Xavier. L'impact des dispositifs médiatiques sur les enfants d'âge préscolaire en situation d'apprentissage avec adulte. Étude d'un cas de simulateur informa- tique dans le contexte d'une exposition scientifique. Thèse
de doctorat en sciences de l'éducation, Université Paris 5 – René Descartes, 2006.

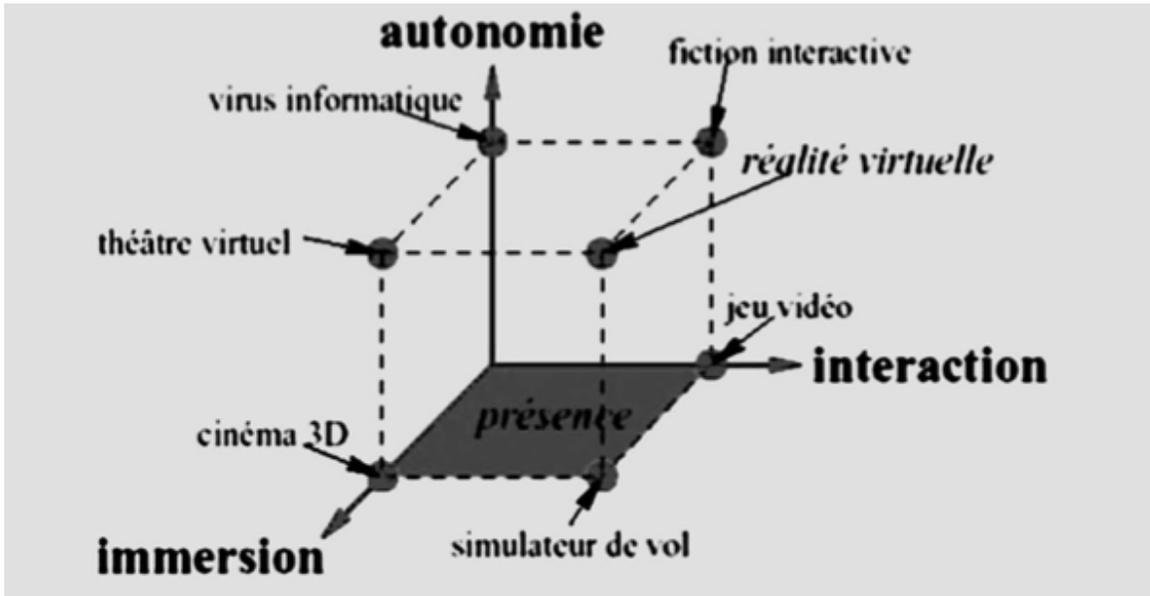

Illustration 1. Présence et autonomie. Tiré de Tisseau et Harrouet, 2003

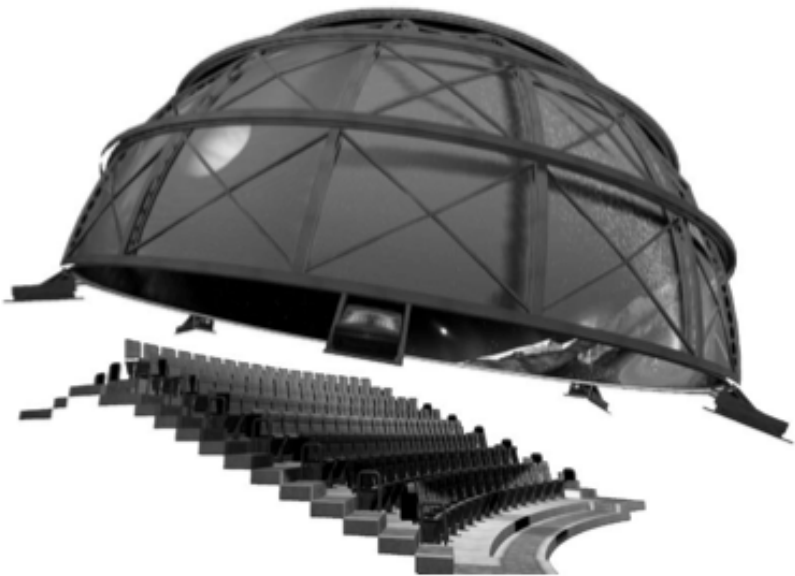

Illustration 2. Configuration d'un planétarium numérique. Modèle architectural d'un planétarium numérique – Evans & Sutherland.

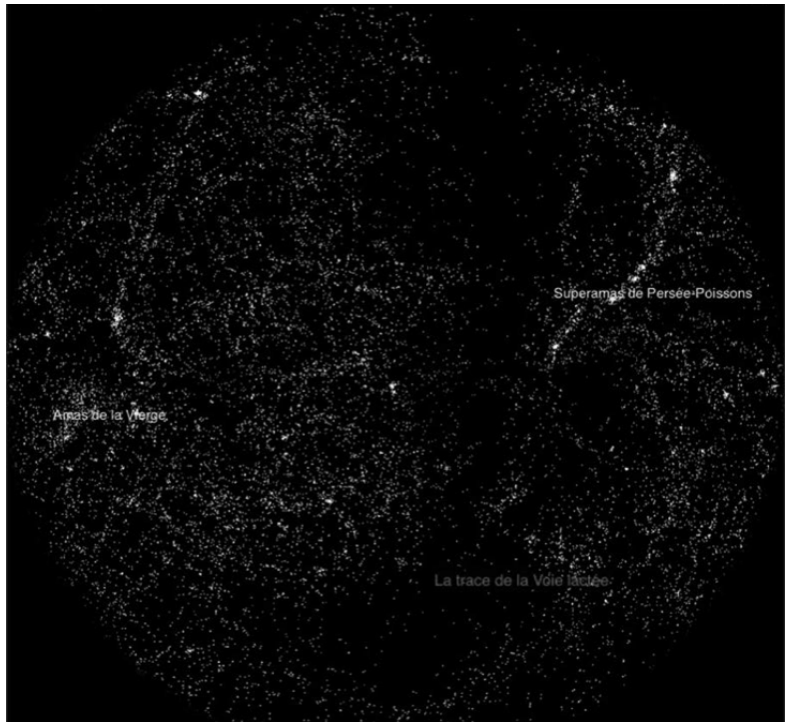

Illustration 3. Représentation de la cartographie galactique en projection 2D sur une voûte. Image tirée d'une séance au Planétarium de Vaulx-en-Velin, janvier 2013.

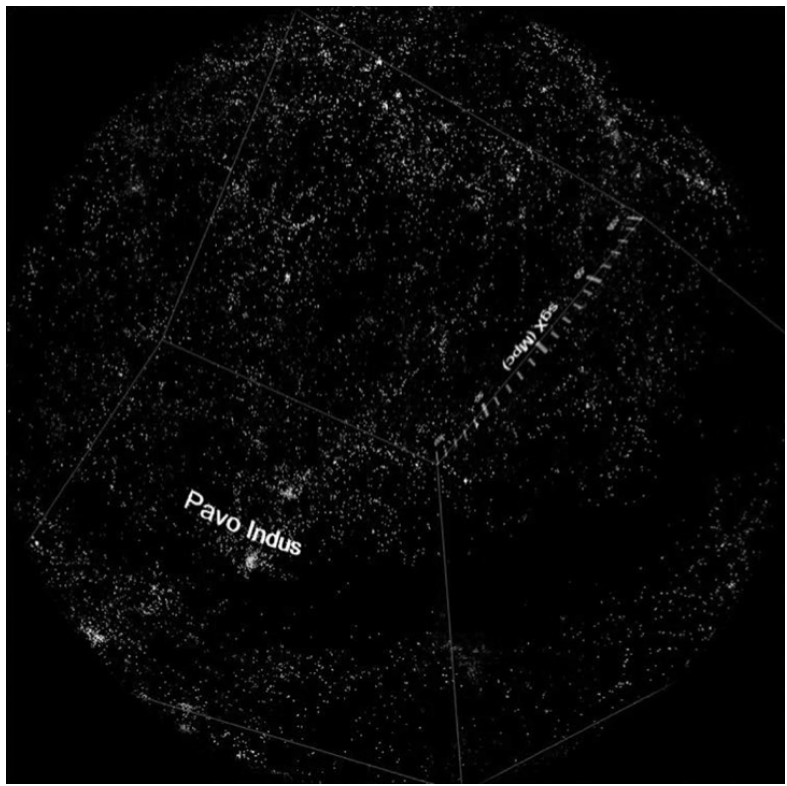

Illustration 4. Représentation de la cartographie galactique en projection 3D dans un rayon de 100 Mégaparsecs. Image tirée d'une séance au Planétarium de Vaulx-en-Velin, janvier 2013.

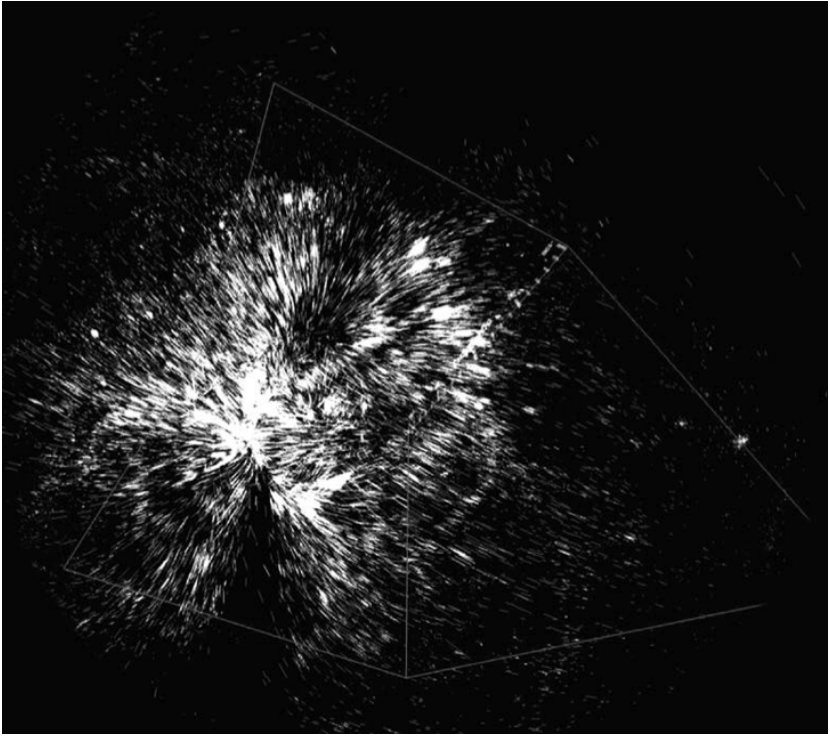
Illustration 5. Représentation des vitesses particulières de chaque galaxie dans un rayon de 100 Mégaparsecs. Image tirée d'une séance au Planétarium de Vaulx-en-Velin, janvier 2013.

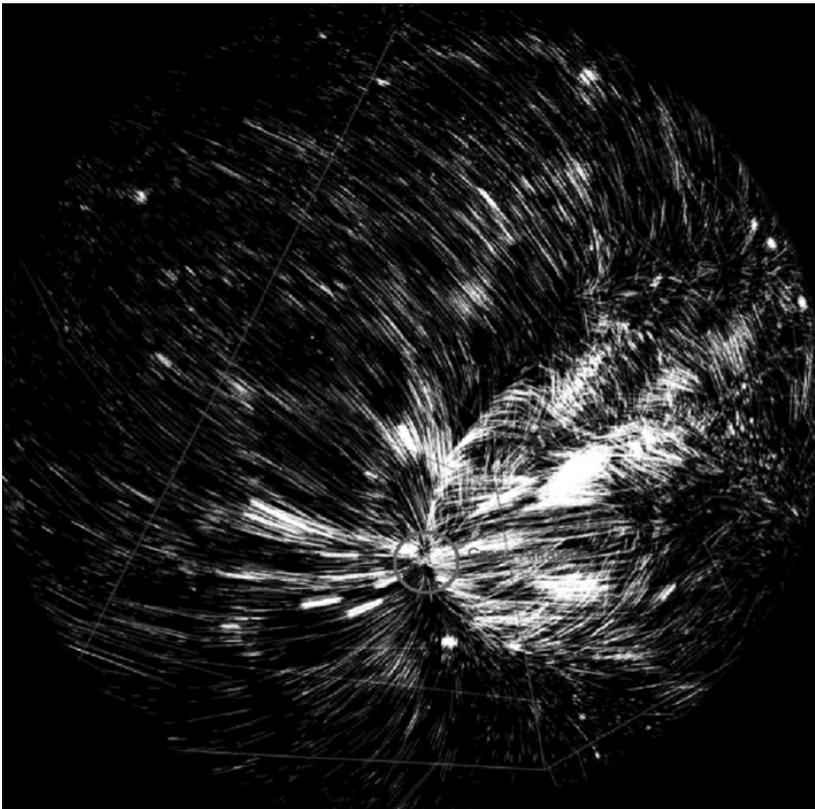
Illustration 6. Position du grand attracteur (cercle rouge). Chaque trait correspond au vecteur vitesse d'une galaxie (rayon de l'Univers 3D : 100 Mégaparsecs). Image tirée d'une séance au Planétarium de Vaulx-en-Velin, janvier 2013.